\RequirePackage[l2tabu, orthodox]{nag}
\PassOptionsToPackage{vvarbb, subscriptcorrection, timesmathacc, slantedGreek}{newtxmath}
\PassOptionsToPackage{theoremfont}{newtx}

\PassOptionsToPackage{unicode, pdfa}{hyperref}
\PassOptionsToPackage{table, hyperref}{xcolor}

\documentclass[english,screen,pbalance,nonacm]{acmart}
\pdfoutput=1 

\usepackage{microtype}

\usepackage{babel}
\usepackage[autostyle=true]{csquotes}

\usepackage{nicematrix}

\NiceMatrixOptions{cell-space-limits = 2pt}

\usepackage{tikz}
\usetikzlibrary{fit}

\usepackage{mathtools, bm}

\usepackage{enumerate}
\usepackage{array}
\usepackage{enumitem}

\usepackage{graphicx}
\usepackage{cleveref}

\DeclareMathOperator{\lcm}{lcm}
\newcommand{\In}{{\mathrm{\,In}}}
\newcommand{\qs}{qs}

\newcommand{\IR}{\mathbb{R}}
\newcommand{\IN}{\mathbb{N}}
\newcommand{\IZ}{\mathbb{Z}}

\newcommand{\IQ}{\mathbb{Q}}
\newcommand{\Qplus}{\mathbb{Q}_{\geq 0}}

\newcommand{\dG}{\mathbb{G}}

\newcommand{\diag}{\mathrm{diag}}

\renewcommand{\leq}{\leqslant}
\renewcommand{\geq}{\geqslant}

\newcommand{\outd}{d^{\scriptscriptstyle{-}}}

\newcommand{\eqdef}{\coloneq}

\newcommand{\zerovec}{{\mathbf{0}}}

\DeclarePairedDelimiterXPP \bigO [1] {\mathrm{O}} () {} {#1}

\newrobustcmd   \mat [1] {\bm{#1}}
\renewrobustcmd \vec [1] {\bm{#1}}
\newrobustcmd \matid  {\mathbf{I}} 

\RequirePackage[ruled,
  vlined,
  linesnumbered,
  inoutnumbered,
  commentsnumbered]{algorithm2e}
\SetKwInOut{Input}{Input}
\SetKwBlock{Initially}{Initially:}{end}
\SetKwBlock{EachRound}{In each round:}{end}
\SetKwComment{Comment}{\(\triangleright\:\)}{}
\SetKwFunction{WR}{write/read}
\SetKwFunction{Return}{return}
\SetKw{Let}{let}
\DontPrintSemicolon

\date{\today}

\setcopyright{cc}
\setcctype{by-nc-sa}

\begin{document}

\title{Know your audience}
\subtitle{Communication model and computability in anonymous networks}

\author{Bernadette Charron-Bost}
\orcid{0009-0007-0132-8138}
\affiliation{%
	\institution{CNRS, DI ENS}
	\institution{École Normale Supérieure}
	\postcode{75005}
	\city{Paris}
	\country{France}}
\email{charron@di.ens.fr}
\author{Patrick Lambein-Monette}
\orcid{0000-0002-9401-8564}
\affiliation{%
	\institution{Unaffiliated}
	\city{Paris}
	\country{France}}
\email{patrick@lambein.name}

\begin{abstract}
	Distributed function computation is the problem,
		for a networked system of~$n$ autonomous agents,
		to collectively compute the value~$f(\omega_1, \ldots, \omega_n)$
		of some input values~$\omega_1, \ldots, \omega_n$,
		each initially private to one agent in the network.
	As is commonplace in the field of distributed computing,
		the question of \emph{which functions~$f$ are computable}
		is exquisitely sensitive to minute model assumptions.
	Here, we study and organize results
		pertaining to distributed function computation
		in anonymous\footnote{%
			When agents are given unique identifiers,
				the question of computability is rather shallow,
				as essentially all functions are computable given sufficient connectivity.%
		} networks, both for the static and the dynamic case,
		under a communication model of directed and synchronous message exchanges,
		but with varying assumptions in the degree of awareness or control
		that a single agent has over its \emph{outneighbors},
		i.e., the set of agents hearing from it in a given round.

	Our main argument is three-fold.
	First, in the \enquote{blind broadcast} model,
		where in each round an agent merely casts out a unique message
		without any knowledge or control over its addressees,
		the computable functions are those
		that only depend on the \emph{set}~$\left\{\omega_1, \ldots, \omega_n\right\}$
		of the input values,
		but not on their multiplicities or relative frequencies in the input.
	Second, in contrast, when we assume either that
		\textbf{a)} in each round, the agents know how many outneighbors they have;
		\textbf{b)} all communications links in the network are bidirectional; or
		\textbf{c)} the agents may address each of their outneighbors individually,
		then the set of computable functions grows to contain all functions
		that depend on the \emph{relative frequencies} of each value in the input --
		such as the \emph{average}~$\frac{\omega_1 + \cdots + \omega_n}{n}$ --
		but \textbf{not} on their \emph{multiplicities} --
		thus, not the \emph{sum}~$\omega_1 + \cdots + \omega_n$.
	Third, however, if one or several agents are distinguished as \emph{leaders},
		or if the cardinality of the network is known,
		then under any of the above three assumptions
		it becomes possible to recover the complete multiset $[\omega_1, \ldots, \omega_n]$
		and thus compute any function of the distributed input
		as long as it is invariant under permutation of its arguments.
	In the case of dynamic networks,
		we also discuss the impact of multiple connectivity assumptions.
\end{abstract}

\maketitle

\section{Introduction}


The aim of many multi-agent systems
	is to get all agents to compute a common value,
	which is a function of the values
	observed or sensed by each agent.
Some typical examples are minimum-finding,
	the computation of the average
	and of the sum of the agent values.
The computation of such functions naturally arises in a wide range of practical situations,
	including sensor networks, distributed optimization, or distributed control in autonomous systems.
Towards this purpose,
	the agents repeatedly alternate
	between internal computations
	and communicating with each other.
Here, our objective is to understand
	the fundamental limitations and capabilities for function computation
	that are inherent to the communication model
	assumed when considering a multi-agent network system.

Our abstract model captures common requirements
	for a variety of different settings,
	including the case of wireless sensor networks.
We consider a networked system
	with a fixed set of agents
	and communication links that may vary over time.
Our basic connectivity assumption is of a \emph{finite diameter},
	i.e., any pair of agents can communicate, possibly indirectly,
	over a period of time that is \emph{uniformly bounded} throughout the execution.
We model agents as automata
	interacting over reliable communication channels.
These automata are \emph{deterministic} --
	in particular, randomization is \emph{not} allowed --
	and identical -- i.e., each agent runs the same local algorithm.

The network is anonymous: agents do not possess unique identifiers,
	and nothing initially distinguishes any two agents
	apart from, possibly, their input value.
In addition, agents have limited or no knowledge of the network.
In particular, they are not assumed to know the size of the network;
	at best, they have an upper bound over it.
They are also unaware of the structure of the network or its diameter.
Concerning the memory of each agent,
	it cannot be bounded and must grow with the size of the network,
	given that the class of functions under consideration
	includes the sum, the average\textellipsis{}
However, for our positive results,
	we will be looking for finite-state solutions:
	if possible, an agent should only use bounded memory in any execution.
We will also be looking for \emph{self-stabilizing} algorithms,
	i.e., algorithms which tolerate arbitrary initializations,
	and for algorithms that tolerate \emph{asynchronous starts}
	(which is obviously the case of self-stabilizing algorithms).

We then consider four classical models for inter-agent communication.
In the lowest-level model
	-- namely, the \emph{simple broadcast} model,
	an agent \enquote{blindly} sends a message,
	without knowing by whom, or by how many,
	this message will be received;
	the content of the message
	is entirely determined by the local state of the agent,
	and is the same for every recipient.
This model can be enriched in two ways,
	either with the feature of \emph{symmetric communications}
	-- an agent~$j$ receives a message from another agent~$i$
	if and only if~$i$ itself receives a message from~$j$ --
	or with \emph{outdegree awareness}
	-- i.e., when an agent broadcasts a message,
	it knows in advance how many other agents will receive this message.
In the latter case, the content of the message is still the same for every recipient,
	but is no longer determined solely by the local state of the sender,
	as it can also depend on its current outdegree.

Symmetric communications arise in many natural systems
	-- such as the popular Hegselmann-Krause model
	used to study the dynamics of opinion formation~\cite{HegselmannK:jasss:2002} --
	and is a basic feature of the pairwise interactions
	of the celebrated \emph{population protocols} model~\cite{AngluinADFP:dc:2006}.
Outdegree awareness is often implicitly assumed
	by distributed algorithms designed for engineering systems,
	such as the credit-recovery algorithm for termination detection~\cite{Mattern:ipl:1989}
	or the Push-Sum algorithm used for decentralized optimization problems~\cite{KempeDG:focs:2003}.

A final communication model is given by \emph{output port awareness},
	in which case each agent is aware of output ports
	corresponding to each communication links.
This amounts to a local output labelling:
	the outgoing links of an agent~$i$ have unique labels,
	and the messages sent by~$i$ over a given link
	may depend on the corresponding label.
In this model, it is no longer the case
	that agent~$i$ sends the same message to each of its neighbors.
We note in passing that this model is only meaningful
	in the context of \emph{static} networks.

We will consider a general definition of what it means
	for a networked system to \enquote{compute} a function:
	given a metric space~$(X, \delta)$
	and a function of arbitrary arity~$f : \bigcup_{n\in \IN_{>0}} \, \Omega^n \rightarrow X$,
	each agent holds an output that must asymptotically converge in~$(X, \delta)$
	to the value taken by~$f$
	when its arguments correspond to the agents' input values.
In particular, when~$X$ is endowed with the discrete metric,
	all agents must eventually settle on the desired value,
	but they are \emph{not} required to become aware when their outputs stabilize.
In contrast, when~$X = \mathbb{R}^k$ is equipped with the Euclidean metric,
	the outputs need only converge towards the desired outcome
	but are not  required to ever stabilize on its exact value;
	this is actually an extremely common metric choice
	in the literature on distributed control.
In all cases, regardless the metric on~$X$, our definition of function computation contains no 
	\emph{termination} requirement: an agent is \emph{not} supposed to return a value that 
	is the desired value exactly, or to within a given precision.

For the simple broadcast model,
	a function~$f$ is computable
	if and only if~$f$ is \emph{set-based},
	i.e., if its value only depends
	on the \emph{set} of its arguments.
Set-based functions include, for example,
	the minimum and the maximum,
	but not the sum or the average.
This result holds for any metric~$\delta$,
	for static and dynamic networks,
	and assuming or not that the network size is known.
The simple gossip algorithm clearly computes the set of input values,
	and thus any set-based function.
For the impossibility result,
	it was shown by Hendrickx and Tsitsiklis
	for an arbitrary network size~\cite{HendrickxT:allerton:2015},
	but the stronger version under the assumption of a known network size
	had been previously shown by Boldi and Vigna~\cite{BoldiV:sirocco:1997}.

In this paper,
	we characterize the functions
	that are computable under the assumptions
	of symmetric communications, output port awareness, or simply outdegree awareness.
In particular, we show that these communication models are actually equivalent
	in terms of function computability.

\paragraph{Summary and contributions.}
First, we provide a general model of computation in anonymous networks with the features described
	above and which encompasses the various communication models that we have just introduced.

Then, we tackle the case of static networks:
	we prove that with either output port awareness, symmetric communications, or  outdegree awareness,
	a function $f$ is computable if and only if $f$ is \emph{frequency-based}, i.e., its value only depends 
	on the set of its  arguments and their \emph{frequencies}.
In particular, computing the average of initial values is possible while it is impossible with the simple broadcast model.
However, computing the sum of initial values remains impossible.
This result holds in any of the three communication models under consideration, for any metric $\delta$,
	and assuming or not that a bound on the network size is known.
Our approach for both positive and negative results exploits the notion of \emph{graph fibration},
	which originated from homotopy theory and has been used first in order to characterize the classes of
	anonymous networks in which leader election is possible~\cite{Angluin:stoc:1980,YamashitaK:wdag:1989,BoldiSVCGS:istcs:1996}.

The impossibility results all use the fundamental \emph{lifting lemma}~\cite{BoldiV:dc:2002}
	stating that all the agents in the same  \emph{fibre},
	i.e., with similar in-neighborhood, have the same behavior
	if they start in the same state.
In fact, our impossibility proofs are a formalization in terms of graph fibration of the argument used in \cite{HendrickxOT:tac:2011}.

The first step for our positive results is the distributed algorithm that computes the \emph{minimal base} of the 
	network~\cite{BoldiV:dc:2002}.
Then we show that the minimal base allows for computing the cardinalities of the fibres up to some common factor.
For that, each agent solves a homogeneous system, which we prove to be  of rank one in each of the three communication models 
	under consideration.
While the latter property clearly holds in the cases of output port awareness and symmetric communications,
	it requires a more sophisticated argument for the model with outdegree awareness,
	which has led us to develop a method \emph{à la} Perron-Frobenius
	for matrices whose diagonal entries may be negative.
The result is a self-stabilizing and finite-state algorithm for computing a frequency-based function 
	in a static network that is linear in time in the number of agents.

As a consequence, when the network size is known or if the network is able
	to appoint a leader, our approach allows for computing
	any \emph{multiset-based} function of the initial values,
	that is to say any function invariant under permutation.
Hence, each of these two assumptions considerably increases the computational power in the case of output port awareness,
	symmetric communications, or outdegree awareness, while they leave it unchanged in the simple broadcast model~\cite{BoldiV:sirocco:1997}.

\Cref{fig:tab} summarizes the computability results for static networks
	in our considered communication models
	and under various assumptions of centralized help
	(knowledge of~$n$ or of a bound over~$n$, presence of a leader).

\begin{table}[htb]
	\caption{Computable functions in static, strongly connected networks of~$n$ anonymous agents}\label{fig:tab}
	\begin{NiceTabular}{r|*{4}{c}}[margin = 2pt]
		\CodeBefore
			\rectanglecolor{blue!15}{2-2}{5-2}
			\rectanglecolor{red!15}{2-3}{3-5}
			\rectanglecolor{green!15}{4-3}{5-5}
		\Body
		\toprule
		\RowStyle{\scshape}
		& \Block[c]{}{simple \\ broadcast}
		& \Block[c]{}{outdegree \\ awareness}
		& \Block[c]{}{symmetric \\ communications}
		& \Block[c]{}{output port \\ awareness} \\[.5em]
		\midrule
		no centralized help
		& \Block[c]{}{set-based \\ \scriptsize Hendrickx \textit{et al.}~\cite{HendrickxOT:tac:2011}}
		& \Block[c]{}{frequency-based \\ {\scriptsize \Cref{thm:computability}, \cref{eq:od}}}
		& \Block[c]{}{frequency-based \\ {\scriptsize \Cref{thm:computability}, \cref{eq:sym}}}
		& \Block[c]{}{frequency-based \\ {\scriptsize \Cref{thm:computability}, \cref{eq:op}}} \\
		a bound over~$n$ is known
		& \Block[c]{}{set-based \\ \scriptsize Boldi \& Vigna~\cite{BoldiV:sirocco:1997}}
		& \Block[c]{}{frequency-based \\ {\scriptsize \Cref{cor:boundn}, \cref{eq:od}}}
		& \Block[c]{}{frequency-based \\ {\scriptsize \Cref{cor:boundn}, \cref{eq:sym}}}
		& \Block[c]{}{frequency-based \\ {\scriptsize \Cref{cor:boundn}, \cref{eq:op}}} \\
		$n$ is known
		& \Block[c]{}{set-based\tabularnote{\footnotesize
				This is in fact for~$n \geq 4$;
					for smaller networks,
					the topology always allows for the recovery
					of the multi-set of input values,
					as Jérémie Chalopin pointed out to us.}
				\\ {\scriptsize Boldi \& Vigna~\cite{BoldiV:sirocco:1997}}}
		& \Block[c]{}{multiset-based \\ {\scriptsize \Cref{cor:n}, \cref{eq:od}}}
		& \Block[c]{}{multiset-based \\ {\scriptsize \Cref{cor:n}, \cref{eq:sym}}}
		& \Block[c]{}{multiset-based \\ {\scriptsize \Cref{cor:n}, \cref{eq:op}}} \\
		one leader
		& \Block[c]{}{set-based\tabularnote{\footnotesize
			Even though Boldi and Vigna~\cite{BoldiV:sirocco:1997}
				do not consider networks with a leader,
				their impossibility argument can be adapted to this case.}
			\\ {\scriptsize Boldi \& Vigna~\cite{BoldiV:sirocco:1997}}}
		& \Block[c]{}{multiset-based \\ {\scriptsize \Cref{cor:leader}, \cref{eq:od}}}
		& \Block[c]{}{multiset-based \\ {\scriptsize \Cref{cor:leader}, \cref{eq:sym}}}
		& \Block[c]{}{multiset-based \\ {\scriptsize \Cref{cor:leader}, \cref{eq:op}}} \\
		\bottomrule
		\CodeAfter
		\tikz \draw [thick, line cap = round] (2-|3) -- (last-|3) (4-|3) -- (4-|last) ;
	\end{NiceTabular}
\end{table}

For dynamic networks, we first observe that the above impossibility results also hold in this  case  since it encompasses
	static networks.
For the positive results, we develop a different method based on the stochastic analysis of consensus algorithms
	derived from statistical physics, namely the \emph{Metropolis} and the \emph{Push-Sum} algorithms.
These algorithms allow for asymptotically computing the average of initial values in dynamic networks.
They are not efficient in time and are not self-stabilizing, but they tolerate asynchronous starts and use no persistent memory.
We provide a concise and streamlined proof of the convergence of the Push-Sum algorithm to the average of the initial values  
	in a dynamic network with a finite diameter.

In the case an upper bound on the network size is available, we obtain the same characterization of 
	computable functions in dynamic networks with symmetric communications
	or outdegree awareness as in the static case, and this characterization holds for any metric,
	in particular for the discrete metric (exact computation).
When no bound on the network size is known,  these algorithms only achieve an approximate computation  
	of the frequencies of the initial values, and thus of  frequency-based functions if they satisfy some 
	continuity property with respect to the frequencies of their arguments.
Typical examples of such functions, which we call \emph{continuous in frequency}, are the average function and
	the \emph{threshold frequency predicates} with a non-rational threshold.
We have thus proven that, in dynamic networks, the frequency-based condition which is necessary for approximate computability,
	is nearly sufficient in the sense that it simply needs to be enriched with continuity in frequency.

The results pertaining to dynamic networks are collected in \Cref{fig:tabdyn}.

\begin{table}[htb]
	\caption{Computable functions in dynamic networks of~$n$ anonymous agents with finite diameter}\label{fig:tabdyn}
	\begin{NiceTabular}{r|*{3}{c}}[margin = 2pt]
		\CodeBefore
			\rectanglecolor{blue!15}{2-2}{5-2}
			\cellcolor{red!15}{2-4,3-3,3-4}
			\cellcolor{green!15}{4-3,4-4,5-4}
		\Body
		\toprule
		\RowStyle{\scshape}
		& \Block[c]{}{simple \\ broadcast}
		& \Block[c]{}{outdegree \\ awareness}
		& \Block[c]{}{symmetric \\ communications}
		\\[.5em]
		\midrule
		no centralized help
		& \Block[c]{}{set-based \\ \scriptsize Hendrickx \textit{et al.}~\cite{HendrickxOT:tac:2011}}
		& \Block[c]{}{{\Large ?}}
		& \Block[c]{}{frequency-based \\ {\scriptsize Di Luna \& Viglietta~\cite{LunaV:disc:2023}}}
		\\ a bound over~$n$ is known
		& \Block[c]{}{set-based \\ \scriptsize Hendrickx \textit{et al.}~\cite{HendrickxOT:tac:2011}}
		& \Block[c]{}{frequency-based \\ {\scriptsize \Cref{cor:boundndyn}}}
		& \Block[c]{}{frequency-based \\ {\scriptsize CB \& LM~\cite{Charron-BostL:sand:2022}}}
		\\ $n$ is known
		& \Block[c]{}{set-based \\ {\scriptsize Hendrickx \textit{et al.}~\cite{HendrickxOT:tac:2011}}}
		& \Block[c]{}{multiset-based \\ {\scriptsize \Cref{cor:ndyn}}}
		& \Block[c]{}{multiset-based \\ {\scriptsize CB \& LM~\cite{Charron-BostL:sand:2022}}}
		\\ one leader
		& \Block[c]{}{set-based \\ {\scriptsize Hendrickx \textit{et al.}~\cite{HendrickxOT:tac:2011}}}
		& \Block[c]{}{{\Large ?}}
		& \Block[c]{}{multiset-based \\ {\scriptsize Di Luna \& Viglietta~\cite{LunaV:focs:2022}}}
		\\ \bottomrule
		\CodeAfter
		\tikz \draw [thick, line cap = round]
			(2-|3) -- (last-|3)
			(4-|3) -- (4-|last)
			(3-|3) -- (3-|4) -- (2-|4)
			(5-|3) -- (5-|4) -- (6-|4);
	\end{NiceTabular}
\end{table}

\paragraph{Related works.}
There is a very large literature on computability in multi-agent systems,
	but most of it focuses on computing functions
	whose values may depend on the network topology,
	and not only on the initial values.
Moreover, a common requirement
	is that all agents become aware that they produce the desired outputs.
We refer the reader to~\cite{Angluin:stoc:1980,MoranW:siamcomp:1993,YamashitaK:tpds:1996,FichR:dc:2003}
	for some fundamental results in this setting.

The biologically-inspired \emph{population protocols} model
	has some common features with our model,
	namely a fixed set of anonymous agents with pairwise interactions\footnote{%
	This corresponds to a dynamic network with symmetric communications and vertices of degree
	zero or one.}
	and no requirement of termination awareness.
However, agents in this model are finite-state,
	and the fairness condition on interactions,
	despite implying that every pair of agents communicate infinitely often,
	does \emph{not} require a bounded dynamic diameter.
A spectacular result is the characterization of the predicates that are computable
	by population protocols: Angluin et al.~\cite{AngluinADFP:dc:2006,AngluinAER:dc:2007} proved that the
	class of computable predicates is exactly the class of predicates definable in Presburger arithmetic.

Closest to our work in the static case are~\cite{BoldiV:sirocco:1997} and~\cite{HendrickxOT:tac:2011}.
The first paper characterizes the class of functions that are computable with simple broadcast or with
	symmetric communications when the network size is known.
The second paper gives an almost characterization of the computable functions in a static network
	with symmetric communications when no bound on the network size is known: 
	similarly to our results in the dynamic case, Hendrickx et al. proved that the  frequency-based  
	condition is sufficient only for an \emph{approximate} computation
	of the frequencies of the initial values.
Hence, we have solved the open question proposed in~\cite{HendrickxOT:tac:2011} of an exact characterization
	of computable functions with symmetric communications, and have extended it to communication models
	with output port awareness and outdegree awareness.

For dynamic networks, our positive result has to be compared with the remarkable algorithm proposed by 
	Di Luna and Viglietta~\cite{LunaV:focs:2022,LunaV:disc:2023} which allows for an exact computation of any 
	frequency-based function in the case of symmetric communications. 
Their algorithm is linear in time, but it uses an infinite number of states  and an infinite bandwidth in each of its executions.
Moreover, it is not self-stabilizing and does not even tolerate asynchronous starts.

\section{The computing model}

\subsection{Networked systems}\label{sec:networks}

We consider a networked system with a fixed and finite set of agents denoted $1,\cdots,n$.
Computation proceeds in  \emph{synchronized rounds}, which are communication closed 
	in the sense that no message received in round $t$ is sent in a round different from~$t$.
In round $t \, (t = 1, 2, \cdots)$, each agent successively (a) sends messages at the beginning of round~$t$,
	(b) receives some messages, and (c) undergoes an internal transition to a new state.
Communications that occur at round $t$ 
	correspond to the (directed) graph 
	$\dG(t)=([ n ] ,E_t)$ where $[ n ]  = \{1, \dots, n\}$ and $(i,j)\in E_t$ if and only if the agent~$j$ receives
	 the message sent by $i$  to~$j$ at round~$t$.
Hence, the agent~$i\in [n]$
	corresponds to the vertex~$i$ in each graph~$\dG(t)$,
	and we will sometimes refer to an agent or a vertex
	as a \emph{node} of the network.
We assume a self-loop at each vertex in each graph~$\dG(t)$,
	since an agent can communicate with itself instantaneously.
The \emph{network} is thus modeled by the \emph{dynamic graph}~$\dG$,
	i.e., the infinite sequence of graphs~$\dG=\left (\dG(t) \right )_{t \geq 1}$,
	with the same set of vertices.

Each agent may possess more or less information about the network it belongs to.
 The knowledge of certain informations thus corresponds to  some constraints on the network,
 	and so to restrict to a non-empty subset of networks, called a \emph{network class}.
Since we consider \emph{anonymous networks}, 
	network classes are assumed to be closed under graph isomorphisms.
As an example, the network class
	where the number of agents is known to be $n$
	is captured by the set of dynamic graphs with $n$ vertices.
Similarly, the class of \emph{symmetric networks}
	corresponds to the set of dynamic graphs with bidirectional edges
	-- that is, at each round~$t$, $(i,j) \in E_t$ if and only if $(j,i) \in E_t$.
We also consider the class of networks~$\dG$ with a finite \emph{dynamic diameter}
	-- that is, there exists some positive integer~$D$
	such that, for every $t \in \IN_{>0}$,
	the graph product\footnote{%
	Recall that the product $G= G_1 \circ G_2$ of two directed graphs $G_1 = (V, E_1)$ and $G_1 = (V, E_2)$ with the
		same set of vertices~$V$ is the directed graph $G = (V, E)$ where 
		$E =  \{ (j,i) \in V^2 : \exists k\in V, (i,k) \in E_1 \wedge (k,j) \in E_2 \} $.}
	$\dG(t) \circ \cdots\circ \dG(t+D-1)$ is the complete graph.
In other words, for every pair of vertices $i, j$ and from every round~$t$,
	there is a \emph{dynamic path} of length at most~$D$ connecting~$i$ to~$j$;
	the smallest such integer~$D$
	is called the \emph{dynamic diameter of} $\dG$.
It measures connectivity \emph{over time} and generalizes the diameter of static graphs.
Note that with a dynamic diameter~$D \geq 2$,
	some intermediate graphs in any period of length~$D$
	may be disconnected (e.g., with only self-loops).

\subsection{Algorithms, communication models, and executions}\label{sec:algocomexec}

An \emph{algorithm}~$\mathcal{A}$ is given by
	a set $\mathcal{Q}$ of local states,
	a subset $\mathcal{Q}_0\subseteq \mathcal{Q}$ of initial states,
	a set of messages~$\mathcal{M}$,
	a sending function,
	and a transition function.

The transition function determines the state after a transition:
	the new state is computed on the basis of the current state
	and the collection of messages that have been received.
That corresponds to a transition function~$\delta : \mathcal{Q} \times \mathcal{M}^{\oplus} \rightarrow \mathcal{Q}$,
	where $\mathcal{M}^{\oplus}$ denotes the set of finite multi-sets over the set~$\mathcal{M}$.

The messages to be sent by an agent
	depend on its current state
	and on its out-neighborhood,
	which can be compactly described by a local output labelling:
	if the outdegree is $\outd$,
	then the output ports are labelled with the numbers $1, \ldots, \outd .$
In this communication model,
	called \emph{output port awareness},
	a sending function is thus of the type~%
	$\sigma : \mathcal{Q} \times \IN_{>0} \to \bigcup_{k \in \IN_{>0}} \mathcal{M}^k$.
If $q$ is the state of an agent and $\outd$ is its outdegree,
	the message sent by the agent 
	on the port labelled by $\ell \in [\,\outd\,]$
	is the $\ell$-th entry of $\sigma(q, \outd) \in \mathcal{M}^{\outd}$,
	denoted $\sigma(q, \outd)[\ell]$.

We can weaken this model
	by considering only the sending functions satisfying
	\begin{equation*}
		\forall q \in \mathcal{Q},
		\forall k \in \IN_{>0},
		\forall \ell, \ell' \in [\,k \,]^2,
		\ \sigma(q,k ) [\ell] = \sigma(q,k ) [\ell'] .
	\end{equation*}
In this communication, model called \emph{outdegree awareness},
	communications are isotropic
	-- a sender sends the same messages to all its recipients --
	and sending functions
	are actually of type~$\sigma : \mathcal{Q} \times \IN_{>0} \to \mathcal{M}$.

A further weakening of the communication model
	consists in requiring messages
	to depend only on the current state of the sender.
In this \emph{simple broadcast} model,
	sending functions satisfy the following graph-invariant property
	\begin{equation*}
		\forall q \in \mathcal{Q},
		\forall k \in \IN_{>0},
		\forall \ell \in [\, k \,],
		\ \sigma(q,k)[\ell] = \sigma(q,1)[1]
	\end{equation*}
	and thus correspond to functions
	of the type~$\sigma : \mathcal{Q} \to \mathcal{M}$.

Hence, these different notions of sending functions
	yield three communication models
	-- namely, simple broadcast,
	outdegree awareness,
	and output port awareness.
In the latter model,
	from an algorithmic point of view,
	we remark that labelling messages with output ports
	is only useful in the context of a static network
	and fixed output port labellings.

This description may be completed with a fourth model,
	called the model of \emph{symmetric communications},
	which corresponds to the restriction of the simple broadcast model
	to the class of networks with bidirectional links.
For this model, in the case of static networks,
	each agent can determine its outdegree
	at the end of the receiving phase in the first round,
	since it is equal to its indegree,
	i.e., the number of messages it has just received.
In other words, symmetric communications implies outdegree awareness
	in the class of static networks.
This is no longer the case with dynamic networks,
	since the in/outdegree in round~$t$
	is not yet available at the time of emission
	in the simple broadcast model,
	and may differ from one round to the next.

An \emph{execution of} an algorithm~$\mathcal{A}$
	in the dynamic graph~$\dG$
	proceeds as follows:
In each round~$t = 1,2,\dots$,
	each agent applies the sending function~$\sigma$
	to generate the message to be sent on each of its output port,
	then it receives the messages sent by
	its incoming neighbors in the graph~$\dG(t)$,
	and finally applies the transition function $\delta$
	to its current state
	and the multi-set of messages it has just received
	to go to a next state.

An execution of~$\mathcal{A}$ in a network with~$n$ agents
	thus corresponds to an infinite sequence
	of \emph{global states} $C^0, C^1,C^2, \cdots$,
	where a global state is defined as a mapping $C: [n] \to \mathcal{Q}$.
The sequence of global states is entirely determined
	by the initial global state~$C^0$
	and the dynamic graph~$\dG$.
In the rest of the paper,
	we adopt the following notation:
	given an execution of~$\mathcal{A}$,
	the value at the end of round~$t$ of any variable $x_i$,
	local to the agent~$i$,
	is denoted by $x_i(t)$,
	and $x_i(0)$ is the initial value of $x_i$ in this execution.

We may consider the more general model
	of executions with \emph{asynchronous starts}~\cite{Charron-BostM:tcs:2019},
	where each agent is activated in an arbitrary round.
Whether the basic network is static or not,
	this execution model
	can be handled by a simple dynamic graph
	with inactive agents being modeled as isolated vertices.
Regarding eventual convergence properties,
	a \emph{self-stabilizing} algorithm~\cite{Dijkstra:cacm:1974}
	-- i.e., an algorithm that works for an arbitrary initialization --
	obviously tolerates asynchronous starts.
In contrast,
	in the self-stabilizing model,
	an agent cannot measure the time elapsed since it started the computation,
	while it can easily do it
	in the execution model with asynchronous starts.
In this sense, self-stabilization is more restrictive
	than tolerance to asynchronous starts.

\subsection{Computability in a metric space}

Let $\Omega$ be a non-empty set
	and let $(X,\delta)$ be a metric space.
Observe that the topology induced by~$\delta$
	is always coarser than the discrete metric $\delta_0$ defined by
	\begin{equation*}
		\delta_0 (x,y) \eqdef \begin{cases*} 0 & if $x = y$ \\ 1 & otherwise. \end{cases*}
	\end{equation*}
If $X = \IR^k$, then we may also consider the Euclidean distance
	\begin{equation*} \delta_2(\vec{x},\vec{y}) \eqdef \sqrt{(x_1 - y_1)^2 + \cdots + (x_k - y_k)^2  } . \end{equation*}

Let $f: \bigcup_{n\in \IN_{>0}} \, \Omega^n \rightarrow X$
	be a function of arbitrary arity,
	and let $\mathcal{A}$ be an algorithm
	with a set of local states of the form~$\Omega \times X$
	and such that its transition function
	does not modify the first state component in $\Omega$.
The first component of the local state of agent~$i$
	is $i$'s input value, and the second one corresponds
	to the value in $X$ of an output  variable denoted $x_i$.

We say that the algorithm $\mathcal{A}$
	\emph{$\delta$-computes the function $f$ in the network class $\mathcal{C}$}
	if, in every execution of $\mathcal{A}$
	with a network $\dG \in \mathcal{C}$ composed of the agents $1, \cdots, n$
	and with the input values $v_1, \cdots, v_n$ in $\Omega$,
	all sequences $(x_i(t))_{t \in \IN}$ converge
	with respect to the distance $\delta$
	to the same value $x^* = f(v_1,\cdots,v_n)$.

The function $f$ is said to be \emph{$\delta$-computable in the network class $\mathcal{C}$} if there exists an algorithm
	that $\delta$-computes $f$ in~$\mathcal{C}$.
Therefore, if~$f$ is $\delta$-computable in a network class~$\mathcal{C}$,
	then it is $\delta$-computable in any
	subclass of~$\mathcal{C}$.
However, if~$f$ is $\delta$-computable in two network classes~$\mathcal{C}_1$
	and~$\mathcal{C}_2 $,
	then~$f$ may be not $\delta$-computable in~$\mathcal{C}_1 \cup \mathcal{C}_2$.

Since the discrete metric~$\delta_0$
	defines the finest topology on~$X$,
	if~$\mathcal{A}$ $\delta_0$-computes a function $f$,
	then it $\delta$-computes~$f$
	for any distance $\delta$ on $X$.
Moreover, there exists a round
	after which all the variables~$x_i$ are equal to~$x^*$,
	and $\mathcal{A}$ is said to \emph{compute $f$ in finite time}.
If $X = \IR^k$ and $\mathcal{A}$ $\delta_2$-computes  $f$ is the Euclidean distance~$ \delta_2$, then  
	$\mathcal{A}$  computes $f$ asymptotically or approximately.

In this paper, we focus on the class of the \emph{frequency-based} functions~\cite{HendrickxOT:tac:2011}
	whose output values only depend on the set of input values and their frequencies.
For a formal definition, let us first introduce some additional definitions.
We say that a function $\nu:  \Omega \rightarrow \Qplus$ is a  \emph{frequency function} 
	if it is positive for a finite set of values and $\sum_{\omega \in \Omega} \nu(\omega) = 1$.
Given a vector $\vec{v}  \in \Omega^n$,
	the \emph{frequency function in} $\vec{v}$, denoted $\nu_{\vec{v}}$,
	is defined by:
	\begin{equation*}
		\nu_{\vec{v}} : \ \omega \in \Omega
			\mapsto \frac{|{\vec{v}}^{-1}(\omega)|}{n} \in \mathbb Q
	\end{equation*}
	where $ |{\vec{v}}^{-1}(\omega)| $ is the multiplicity of the value $\omega$ in $\vec{v}$.

Conversely, for every frequency function~$\nu$ on~$\Omega$, there exist vectors
	whose frequency functions are equal to~$\nu$:
For instance,  let us consider a total ordering $\omega_1, \cdots, \omega_{\ell}$ on $\nu$'s support,
	irreducible representants $p_k/q_k$ of the positive rational numbers $ \nu(\omega_k)$,
	and the vector $\langle \nu \rangle$ where values in $\nu$'s support occur in order 
	$\omega_1, \cdots, \omega_{\ell}$ and with multiplicities~$p_1q/q_1, \cdots , p_{\ell} q/q_{\ell}$,
	 where $q = \lcm(q_1,\cdots,q_{\ell})$.
This vector of size $q$ will be denoted by $\langle \nu \rangle$ in the sequel.%
\footnote{Actually, the vector $\langle \nu \rangle$
	depends on the total ordering on $\nu$'s support
	and is defined up to some permutation.}

The vectors with the same frequency function $\nu$
	are said to be $\nu$-\emph{frequenced} and \emph{equivalent in frequency}.
	A function~$f: \bigcup_{n \in \IN_{>0}} \Omega^n \rightarrow X$
	is then \emph{frequency-based}
	if~$f$ takes the same value on all vectors equivalent in frequency,
	that is,
	\begin{equation*}
		\forall \vec{v}, \vec{w} \in \Omega^n\times\Omega^m, \ \nu_{\vec{v}} = \nu_{\vec{w}} \implies f(\vec{v})=f(\vec{w}) \, .
	\end{equation*}
Typically, the average function,
	whose value for a vector $\vec{v} \in \IR^n$ is denoted
	$\overline{\vec{v}} = \frac{\sum_{i \in [n]} v_i}{n}$,
	is frequency-based.

We will similarly consider the class of \emph{set-based} functions,
	which take the same value
	on all vectors with the same support,
	and the class of \emph{multiset-based} functions
	-- also referred to as \enquote{symmetric} functions --
	which take the same value on all vectors
	that are permutations of one another.
The three classes clearly obey the inclusion:
	\begin{equation*}
		\text{set-based} \subsetneq \text{frequency-based} \subsetneq \text{multiset-based} \, ;
	\end{equation*}
	examples of functions of each class include, respectively,
	the maximum, the average, and the sum
	of the entries of a vector.

 \section{Graph fibrations}\label{sec:fibration}

A (directed, multi-)graph~$G$
	is composed of a non-empty set of vertices~$V_G = [n_G]$,
	where~$[ n_G ] \eqdef \{1, \dots, n_G\}$
	and a set~$E_G$ of edges
	defined by two functions~$s_G,t_G: E_G \rightarrow V_G$,
	which specify the source and the target vertices of each edge.

A \emph{valued graph} in~$\Omega$
	is a graph~$G$
	together with a function $\vec{v}: V_G \rightarrow \Omega$,
	called a \emph{valuation}.
A \emph{colored graph}, with a set of colors $\mathcal{C}$,
	is a graph $G$ together with a coloring function
	$c: E_G \rightarrow \mathcal{C}$.
In the case where $\Omega$ and $\mathcal{C}$ are singletons,
	the introduction of values and colors makes no difference,
	i.e., vertices are actually not valued and edges not colored.

Outdegree awareness and output port awareness
	then respectively correspond to a valuation and a coloring
	of any arbitrary graph~$G$.
The valued graph, where each vertex~$i$ of~$G$
	is valued with its outdegree~$\outd_i$, is denoted by~$G_{od}$.
Similarly, $G_{op}$ denotes any colored graph
	resulting from a coloring of edges in~$G$
	with local output port labellings.

Our first results with static networks
	rely on the notion of \emph{fibration},
	which originates in homotopy theory, and has been developed
	in order to characterize the classes of static anonymous networks
	in which leader election is solvable~\cite{YamashitaK:tpds:1996,BoldiSVCGS:istcs:1996}
	and functions are computable~\cite{YamashitaK:mst:1996,BoldiV:sirocco:1997}.
	
A \emph{graph morphism}~$\varphi: G \to H$
	is a pair of functions~$\varphi_V: V_G \to V_H$
	and~$\varphi_E: E_G \to E_H$ that commute with the source and target functions,
	i.e., $s_H \circ \varphi_E = \varphi_V \circ  s_G$.
In the valued and colored cases,
	we also require~$\vec{v}_G = \vec{v}_H \circ \varphi_V$
	if~$\vec{v}_G$ and~$\vec{v}_H$
	are valuations of the graphs~$G$ and $H$,
	and $c_G = c_H \circ \varphi_E$
	if $c_G$ and $c_H$ are colors of $G$ and $H$.
When no confusion may arise, subscripts will be dropped.
A \emph{graph isomorphism} is a graph morphism
	that is additionally bijective,
	i.e., both functions~$\varphi_V, \varphi_E$ are bijective.

A \emph{fibration} between (valued, colored) graphs~$G$ and~$B$
	is a morphism $\varphi: G \to B$
	such that for every edge $e \in E_B$
	and for each $i \in V_G$ with $\varphi (i) = t(e)$,
	there exists a \emph{unique} edge $\widetilde{e}^{\,i}$
	verifying $\varphi(\widetilde{e}^{\,i}) = e$
	and~$t(\widetilde{e}^{\,i}) = i$.
We restrict fibrations to be \emph{epimorphisms} --
	that is, $\varphi_V$ and $\varphi_E$ are both surjective.
Under this condition, fibrations preserve strong connectivity.
All graph isomorphisms are fibrations,
	but the converse is generally false.

If $\varphi: G \to B$ is a fibration,
	then $B$ is called the \emph{base} of the fibration.
The \emph{fibre} over a vertex $i$ of the base $B$
	is the set of vertices in $G$
	that are mapped to~$i$,
	and is denoted by $ \varphi^{-1}(i)$.
A fibration $\varphi: G \to B$
	induces an equivalence relation between the vertices of $G$,
	whose classes are precisely the fibres of $\varphi$:
	when two vertices $j$ and $k$
	are in the same fibre,
	they have similar in-neighborhoods --
	that is, there is a bijective correspondence
	between the egres of~$G$ edges coming to $j$
	and those coming to $k$.

\subsection{Lifting lemma}

Impossibility results are based on the fundamental \emph{Lifting lemma}~\cite[Lemma~2]{BoldiV:dc:2002},
	which relates the behaviors of the same algorithm
	on two different networks.
For a formal statement of this lemma,
	we first observe that the notion of execution
	of an algorithm defined in \Cref{sec:algocomexec}
	for communication graphs formed with
	simple edge graphs naturally extends to multi-graphs.
This extension works in the broadcast model
	as well as in the communication models with outdegree awareness
	and output port awareness.
Then we introduce some additional notation:
	if~$\varphi: G \to B$ is a fibration
	and~$C: V_B \to \mathcal{Q}$
	is a global state of the vertices in~$B$,
	then we obtain a global state~$C^{\varphi}$ of the vertices in~$G$
	by copying the state of a vertex of~$B$ fibrewise.
Formally, we let
	\begin{equation*}
		\forall i \in V_G, \ C^{\varphi}_i \eqdef C_{\varphi(i)} .
	\end{equation*}
Similarly, any valuation~$\vec{v}$ of the vertices in~$B$
	can be lifted into the valuation~$\vec{v}^{\varphi}$ of the vertices in~$G$.
From the restriction of the function~$f$ to~$\Omega^{n_G}$,
	we thus define the $n_B$-arity function~$f^{\varphi}$
	as~$f^{\varphi} (\vec{v}) \eqdef f(\vec{v}^{\varphi})$.
We now recall the statement of the Lifting lemma.

\begin{lemma}[Lifting lemma]\label{lem:lifting}
Let $\varphi : G \to B $ be a fibration.
Then, for every algorithm $\mathcal{A}$
	and every computation~$C_0, C_1, \cdots$ of $\mathcal{A}$ on~$B$,
	the sequence $(C_0)^{\varphi}, (C_1)^{\varphi}, \cdots$
	is a computation of~$\mathcal{A}$ on~$G$.
\end{lemma}

By unicity of the limit (with respect to any distance $\delta$),
	the Lifting lemma imposes strong constraints
	on the type of computable functions in a class of networks
	that can be \enquote{collapsed} by fibration
	onto other ones in this class.

\begin{lemma}\label{lem:compbase}
Let $\varphi : G \to B $ be a fibration.
If a function~$f$
	is $\delta$-computed by some algorithm~$\mathcal{A}$
	on both graphs~$G$ and~$B$,
	then $f^{\varphi} = f$.
\end{lemma}

In particular, if the function~$f$ is $\delta$-computable
	in a class that contains both~$G$ and~$B$,
	then the premise of \Cref{lem:compbase} hold and~$f^{\varphi} = f$.
By applying \Cref{lem:compbase}
	to fibrations that are graph isomorphisms,
	and because the network classes under consideration
	are closed under graph isomorphisms,
	we obtain that the value of a computable function
	only depends on the \emph{multi-set} of its arguments,
	and this in any network class considered in this work
	-- that is, only symmetric functions are computable in our model.

\begin{lemma}\label{lem:perm}
	If a function~$f$
		is $\delta$-computed by some algorithm $\mathcal{A}$
		in an anonymous network class,
		then~$f$ is a multiset-based function.
\end{lemma}

Thus, the arguments of a computable function~$f$
	are actually multi-sets in $\Omega^{\oplus}$.
If the support of a vector $\vec{v}\in \IR^n$
	is the set~$\{\omega_1, \cdots, \omega_k \}$
	and if~$\mu_{\ell}$
	denotes the multiplicity of $\omega_{\ell} $ in $\vec{v}$,
	then we have
	\begin{equation*}
		f(\vec{v}) = f (\omega_1^{\mu_1}, \cdots, \omega_{k}^{ \mu_k}),
	\end{equation*}
	where $\omega_{\ell}^{\mu_{\ell}}$ denotes
	the sequence $\omega_{\ell}, \cdots, \omega_{\ell}$,
	of length $\mu_{\ell}$.

\subsection{Minimum base}

A graph~$G$ is said to be \emph{fibration prime}
	if every fibration from $G$ is an isomorphism --
	that is, if~$G$ cannot be collapsed onto a smaller graph by a fibration.
Every graph $G$ has exactly one fibration prime basis (up to some isomorphism),
	called the \emph{minimum base} of $G$.

Boldi and Vigna~\cite{BoldiV:dc:2002}
	constructed a self-stabilizing algorithm
	which distributively computes, in finite time,
	the minimum base of the underlying graph of a static and strongly connected network.
More precisely, when running this algorithm over the static network~$G$,
	each agent builds a graph at the end of each round
	that, from round~$n + D$ onwards,
	is guaranteed to be the minimum base of $G$,
	where $n$ is the number of vertices in~$G$ and~$D$ is its diameter.
Then they proposed a finite-state variant of this algorithm
	that fits our model
	and stabilizes with an overhead of~$\bigO{D \log D}$ rounds
	when compared to the infinite-state version.

\section{Computability in static networks} \label{sec:staticnet}

In this section, we establish our main theorem for static networks, namely that a function of arbitrary arity
	is computable if and only if its output value only depends on the frequencies of 
	input values.
This result holds whatever the distance is and even when agents only know their outdegrees.

\begin{theorem}\label{thm:computability}
Let $(X,\delta)$ be a metric space.
In any of the three communication models
	with either output port awareness, symmetric communications, or outdegree awareness,
	and for any function $f: \bigcup_{n \in \IN_{>0}} \Omega^n \to X$,
	the following assertions are equivalent:
	\begin{enumerate}[label=(\roman*)]
		\item $f$ is frequency-based;
		\item $f$ is $\delta$-computable in the class of static strongly connected networks.
	\end{enumerate}
\end{theorem}

\subsection{Proof of the impossibility result}\label{sec:impoproof}

The main argument in the proof of the impossibility result in \Cref{thm:computability}
	already appears in~\cite{BoldiV:sirocco:1997, HendrickxOT:tac:2011}.
We recall it in order to make the paper self-contained and to show that the result holds whatever the distance is
	and in any of the three communication models under consideration. 

\noindent \emph{Proof of $(ii) \Rightarrow (i)$}.\\
Let $\vec{v}$ and $\vec{w}$ be two $\Omega$-vectors, of respective lengths $n$ and $m$,
	with the same frequency functions $\nu_{\vec{v}} = \nu_{\vec{w}}$.
They share the same support~$\{ \omega_1, \cdots, \omega_{\ell}\} $,
	and we let $\nu_{\vec{v}}(\omega_k) = \frac{p_k}{q_k}$
	with $\gcd(p_k, q_k) = 1$.
Then the integer $p = \lcm (q_1,\cdots,q_{\ell})$ divides both $n$ and $m$.

Let us now consider the bidirectional rings $R^p$, $R^p_{op}$, and $R^p_{od}$, of size $p$,
	with or without output port awareness and outdegree awareness.
The mapping $ i \in [\, n \,] \mapsto  i  \mod p \in  [\, p \,] $
	induces a  fibration $\varphi : R^{n} \rightarrow R^p $.
This fibration preserves both the deterministic coloration of the outgoing links and the outdegree valuation of the vertices.
Similarly, we define a  fibration $\psi : R^{m} \rightarrow R^p $, which is also a fibration of the colored and valued
	rings $R^m_{op}$ and $R^m_{od}$.

Let $\mathcal{A}$ be an algorithm that $\delta$-computes a function $f$ in the class of static strongly connected networks,
	with or without output port or outdegree awareness.
In particular, $\mathcal{A}$ $\delta$-computes $f$ on $R^n$ (resp. $R^n_{op}$ and $R^n_{od}$).
\Cref{lem:compbase} shows that $\mathcal{A}$
	$\delta$-computes the function $f^{\varphi}$ on $R^p$
	(resp. $R^p_{op}$ and $R^p_{od}$).
Similarly, 	$\mathcal{A}$  $\delta$-computes the function $f^{\psi}$ on $R^p$  
	(resp. $R^p_{op}$ and $R^p_{od}$).
We thus obtain that $f^{\varphi} = f^{\psi} $, that is
	for every vector $\vec{u} \in \Omega^p$,
	$f(\vec{u}^{\varphi}) = f(\vec{u}^{\psi})$.

The vector $\vec{u} = (\omega_1^{p\, p_1/q_1}, \cdots, \omega_k^{ p\, p_{\ell}/q_{\ell}} ) = \langle \nu_{\vec{v}} \rangle$ is of size~$p$.
Then $\vec{u}^{\varphi}$ and $\vec{u}^{\psi}$ are respectively obtained
	by permuting the entries of $\vec{v}$ and~$\vec{w}$.
\Cref{lem:perm} shows that $f(\vec{v}) = f(\vec{u}^{\varphi}) $ and $f(\vec{w}) = f(\vec{u}^{\psi}) $.
The equality $f^{\varphi} = f^{\psi}$
	leads to~$f (\vec{v}) = f(\vec{w})$ as required.

\subsection{Positive result with outdegree awareness}

We now devise an algorithm that computes any frequency-based function~$f$
	in the class of static strongly connected networks with outdegree awareness in linear time.
Since the discrete metric~$\delta_0$ defines the finest topology on~$X$, our algorithm
	also $\delta$-computes~$f$ for any distance $\delta$ on $X$.
For that, we use the self-stabilizing algorithm of Boldi and Vigna~\cite{BoldiV:dc:2002}
	that constructs distributively the minimum base of the underlying network~$G$.
Their algorithm relies firstly on the inductive construction by each agent of its \emph{view} in the graph~$G$, 
	and secondly on a partial function $\mathcal{B}$ that allows an agent to extract
	from a truncated and possibly incorrect version of its view
	a candidate for the minimum base of~$G$.
More precisely, in each round~$t$, each agent $i$ builds an in-tree~$T_i^t$ and a multi-graph $\mathcal{B}(T_i^t)$
	that is guaranteed to be the minimum base of the graph~$G$ (up to some isomorphism) from round $n+D$,
	where $n$ is the number of vertices in~$G$ and~$D$ is its diameter.
The algorithm is then made finite-state
	with a loss of less than $D \log(1+D)$ rounds in the stabilization time.
Moreover, it straightforwardly adapts to any valued version of~$G$.

\noindent \emph{Proof of $(i) \Rightarrow (ii)$ in the case of outdegree awareness}.\\
Let  $G_{v, \outd} $ be a double-valued  graph of size $n$ and of finite diameter~$D$.
The first valuation $\vec{v}$
	forms a vector in $\Omega^n$;
	we let~$S$ denote $\vec{v}$'s support.
The second valuation $\outd \in \IN^n$
	is the valuation of the vertices with their outdegrees.
As above explained, at each round $t \geq n +D +  D \log(1+D)$ of Boldi and Vigna's algorithm
	in the graph~$G_{\vec{v}, \outd}$,
	each agent builds the minimum base~$B_{\vec{w} ,\, b}$ of~$G_{\vec{v}, \outd}$.
Let $ \varphi : G_{\vec{v}, \outd} \to B_{\vec{w} ,\, b}$
	be the corresponding fibration
	and let $m = \left|V_B \right|$.

All the vertices of a fibre share the same entry in $\vec{v}$ and the same outdegree:\footnote{%
	Observe that $b_i$ may be different from the outdegree of the vertex $i$ in the graph~$B$.}
	\begin{equation*}
		\forall i \in V_B, \, \forall k \in \varphi^{-1} (i), \ \ w_i = v_k \ \mbox{ and } \ b_i = \outd_k .
	\end{equation*}
Clearly, the $m$ fibres satisfy the following equalities:
	\begin{equation}\label{eq:od}
		b_i \left | \varphi^{-1} (i) \right | = \sum_{j\in V_B} d_{i ,j}  \left | \varphi^{-1} (j) \right | ,
	\end{equation}
	where $d_{i , j} $ denotes the number of edges in $B$ from $i$ to $j$.
The minimum base~$B_{\vec{w},\,b}$
	entirely determines the square matrix~$\mat{M}$
	of size~$m$ defined by
	\begin{equation*}
		M_{i,j} =
			\begin{cases*}
				d_{i, j}       & if $i \neq j$ \\
				d_{i, i} - b_i & if $i = j$.
			\end{cases*}
	\end{equation*}
Let us now consider the linear system $ \mat{M}\, \vec{z} = \zerovec $
	where $\zerovec$ is the zero vector of size $m$.
Observe that the system $\mat{M}\, \vec{z} = \zerovec $
	has a positive solution $\vec{z}$
	whose entries are given by $z_i = \left | \varphi^{-1} (i) \right |$.

Define the matrix $\mat{P} \eqdef \mat{M} + \alpha\, \matid$,
	where $\matid$ is the identity matrix
	and $\alpha$ is an arbitrary real number satisfying
	\begin{equation*}
		\alpha > - \min_{1 \leq i \leq m}  M_{i, i} .
	\end{equation*}
The matrix $\mat{P}$ is non-negative,
	and its diagonal entries are all positive.
Hence, there is a self-loop at each vertex of the associated graph $G_{\mat{P}}$ associated\footnote{%
	See \Cref{subsec:pushsum:preliminaries} for a definition of the associated graph.}
	to $\mat{P}$.
Moreover, $G_{\mat{P}}$ coincides with the support of the graph~$B_{\vec{w},b}$, except possibly for the self-loops.
Therefore, this graph is strongly connected,
	i.e., the matrix $\mat{P}$ is irreducible.

The Perron-Frobenius theorem then applies to the matrix~$\mat{P}$:
	the spectral radius $\varrho_{\scriptscriptstyle \mat{P}}$ of $\mat{P}$
	is an eigenvalue of $\mat{P}$
	of algebraic and geometric multiplicity one,
	and associated with an eigenvector $\vec{x}$ with positive entries.
Hence, $\lambda = \varrho_{\scriptscriptstyle \mat{P}} -\alpha$ is a real eigenvalue of $\mat{M}$,
	and every (complex) eigenvalue of $\mat{M}$ other than $\lambda$
	has a real part less than $\lambda$.
The eigenvalue~$\lambda$ has algebraic and geometric multiplicity one with the positive eigenvector $\vec{x}$.

If $\vec{y} \in \big (\IR_{_{\geq 0}}\big)^m$
	is a non-zero eigenvector of the matrix~$\mat{M}$
	for a real eigenvalue $\mu$,
	then $\vec{y}$ is also an eigenvector of the matrix~$\mat{P}$
	for the real eigenvalue $\mu + \alpha$.
As just shown, we have:
	\begin{equation*} \mu + \alpha \leq \varrho_{\scriptscriptstyle \mat{P}} . \end{equation*}
Let $y_{i}$ be a non-zero entry of the vector~$\vec{y}$.
Since $\mat{P}$ is non-negative, the $i$-th entry of $\mat{P} \, \vec{y}$ satisfies:
	\begin{equation*} (\mat{P} \vec{y})_{i}  = (\mu+ \alpha) \, y_{i} \geq P_{i, i } \, y_{i} \end{equation*}
	which implies that $\mu + \alpha > 0$.
We now show that $\vec{y}$ is  a positive vector.
For that, we use the strong connectivity of the graph $G_{\mat{P}}$,
	and prove by finite induction over $\ell$, $0 \leq \ell \leq m-1$,
	that the entry $y_j$ is positive
	whenever $j$ is at distance $\ell$ from $i$ in $G_{\mat{P}}$.
	\begin{enumerate}
		\item The base case $\ell=0$ is by definition of $i$.
		\item Inductive step: let $\ell\geq 1$
			and assume that the entries of the vector~$\vec{y}$
			for all vertices at distance $\ell-1$ from~$i$ are positive.
			The vertex $j$ has an outgoing neighbor~$k$ at distance $\ell-1$ from~$i$,
			and the $j$-th entry of $\mat{P} \, \vec{y}$ satisfies:
			\begin{equation*} (\mat{P} \, \vec{y})_{j}  = (\mu+ \alpha) \, y_{j} \geq P_{j , k } \ y_{ k } .\end{equation*}
			Since both $P_{j , k} $ and $\mu + \alpha$ are positive,
			the inductive assumption implies that $y_j$ is positive.
	\end{enumerate}

Hence, we can choose a positive real number $\varepsilon$
	small enough in order to have the componentwise inequality:
	\begin{equation*}\varepsilon \, \vec{x} \leq  \vec{y} .\end{equation*}
Since the matrix $\mat{P}$ is non-negative,
	we obtain the following inequality for every positive integer $k$:
	\begin{equation*}
		\varepsilon \, \mat{P}^k (\vec{x})
	= \varepsilon \, (\varrho_{\scriptscriptstyle \mat{P}})^k \vec{x} \leq \mat{P}^k (\vec{y})
	= (\mu+ \alpha )^k \vec{y} .
\end{equation*}
It follows that $\varrho_{\scriptscriptstyle \mat{P}} \leq \mu+ \alpha $,
and thus $\mu = \varrho_{\scriptscriptstyle \mat{P}} - \alpha = \lambda$.

Since the system $\mat{M}\, \vec{z} = \zerovec$ has a positive solution
	whose entries are equal to the cardinalities of the fibres,
	we deduce that $\lambda = 0$,
	and the set of solutions of this system, namely $\ker \mat{M}$,
	is a subspace of dimension one.

In a second step of the algorithm,
	using Gaussian elimination over the Euclidean ring $\IZ$ (see e.g.,~\cite{Jacobson:1980}),
	each agent computes a positive integer vector $\vec{z} \in \IN^m$
	whose all entries are coprime and such that $\ker \mat{M} = \IR \, \vec{z}$.
Subsequently, the agent computes $f(\widetilde{\vec{v}})$
	where $\widetilde{\vec{v}} \in \Omega^{\,p}$ is a vector of size $p \eqdef \sum_{i =1}^m z_{i}$
	and where each value $v_i$ in $S$ occurs with multiplicity~$z_{i}$.
Since we have just proved that
	there exists a positive integer $k$ such that
	\begin{equation}\label{eq:cardfibre}
	 \forall i \in V_B, \ \ \left | \varphi^{-1} (i) \right | = k \, z_i,
	 \end{equation}
	and $f$ is a frequency-based function,
	each agent actually outputs the value $f(\widetilde{\vec{v}}) = f(\vec{v})$.

In each round $t$, each agent~$i$
	builds the finite tree $T_i^t$
	and the (valued) multi-graph $\mathcal{B}(T^t_i)$,
	and then applies the Gaussian elimination method
	over the Euclidean ring
	to solve the linear system corresponding to $\mathcal{B}(T^t_i)$.
The above recalled result by Boldi and Vigna~\cite{BoldiV:dc:2002} on the graphs $\mathcal{B}(T^t_i)$
	implies that, in this way,
	each agent computes the value~$f(\vec{v})$
	no later than in round~$n+D$.

\subsection{Positive results with output port awareness and symmetric graphs}

Each agent can easily retrieve its outdegree in a bidirectional network as well as when it is output port aware.
The above algorithm with a preliminary phase of outdegree calculation thus allows agents to compute 
	any frequency-based function in the  models with symmetric communications or with output port awareness.
We now present two variants of the algorithm for symmetric communications and with output port awareness accordingly,
	which  directly compute frequency-based functions
	without pre-calculation of the outdegrees, leading to  linear systems  that can be easily solved  without  the use of Gaussian elimination.

\paragraph{Output port awareness.} With output port awareness,
	any fibration is actually a \emph{covering},
	i.e., for any pair of vertices $i$ and~$j$ in the same fibre,
	the outgoing edges of~$i$ and~$j$
	are in one-to-one correspondence.
This local isomorphism property
	gives a bijective correspondence
	between the whole neighborhoods of two vertices in the same fibre
	and, as a result,
	the cardinality of all fibres is the same (see e.g.,~\cite{BoldiV:dm:2002}).
In the case of output port awareness,
	\cref{eq:od} is thus replaced by:
	\begin{equation}\label{eq:op}
	\left | \varphi^{-1} (i) \right | =   \left | \varphi^{-1} (j) \right | .
	\end{equation}
Each agent builds the multi-graphs $\mathcal{B}(T_i^t)$
	which are eventually equal to the minimum base~$B_{\vec{w}}$
	of the (colored and valued) graph $G_{\vec{v}}$.
If $p$ denotes the common cardinality of all the fibres,
	then $\vec{v} = (w_1^{p}, \cdots, w_m^{p})$.
It follows that $f(\vec{v}) = f(\vec{w})$
	since~$f$ is a frequency-based function.
In this way, each agent can thus directly compute
	the value $f(\vec{v})$
	from the construction of the valued multi-graph~$B_{\vec{w}}$.

\paragraph{Symmetric communications.}
If the network $G$ is bidirectional,
	then for any fibration $\varphi \, : \, G  \to B$, we have:
	\begin{equation}\label{eq:sym}
		d_{i,j} \, |\varphi^{-1}(j)| = d_{j,i} \, |\varphi^{-1}(i)|
	\end{equation}
	where $d_{i,j}$ denotes the number of $i \to j$ edges
	in the multi-graph $B$.

Let $m$ the number of vertices in the graph~$B$.
Up to some permutation of the vertices in $B$,
	we may assume that none of the degrees
	$d_{1,2}, \cdots,  d_{m-1,m }$ is zero,
	since~$B$ is strongly connected.
Hence, the cardinalities of the fibres
	form a solution of a linear system
	of $m$ equations and $m$ variables
	whose set of solutions is thus of dimension at least one.
Moreover, the positive integer vector $\vec{z}$ defined by:
	\begin{equation*}
		z_i \eqdef
		\begin{cases*}
			d_{1, 2} \times \cdots \times d_{ m-1, m }
				& if $i = 1$ \\
			\frac{d_{2, 1}  \times \, \cdots \,\times d_{i, i-1}}{d_{1, 2} \times \, \cdots \, \times d_{ i-1, i }} \, z_1
				& if $i \neq 1$
		\end{cases*}
	\end{equation*}
	is obviously a basis of the solution set.
Consequently, if $ \widetilde{\vec{v}} $ denotes
	a vector with the same support as $\vec{v}$
	and where each value $v_i \in S$
	occurs with multiplicity $z_i$,
	then $f(\widetilde{\vec{v}}) = f(\vec{v})$,
	since the function~$f$ is frequency-based.
\cref{eq:sym} thus yields an algorithm
	that directly computes the function~$f$
	from the construction of the minimum base $B_{\vec{w}}$
	in the case of a bidirectional network.

\subsection{Computing with knowledge on the network size}

As explained in Section~\ref{sec:networks}, computability of a function when the number of agents is known 
	means that for every positive integer~$n$, there exists an algorithm which computes the function in 
	the class ${\mathcal C}^s_n$ of static strongly connected networks with $n$ agents.
Similarly, the function is computable when an upper bound on the network size is known 
	if for every positive integer~$N$, there is an algorithm which computes the function in the network class 
	${\mathcal C}^s_{\leq N} = {\mathcal C}^s_1\cup\cdots\cup {\mathcal C}^s_N$.

A  refinement of the proof in Section~\ref{sec:impoproof} shows that the impossibility result 
	still holds when a bound on the size of the network is known.
With the above positive results, we thus obtain the following corollary.
	
\begin{corollary}\label{cor:boundn}
Let $(X,\delta)$ be a metric space.
In any of the three communication models with either output port awareness, symmetric communications, 
	or outdegree awareness, and for any function $f: \bigcup_{n \in \IN_{>0}} \Omega^n \to X$,
	the following assertions are equivalent:
	\begin{enumerate}[label=(\roman*)]
		\item $f$ is frequency-based;
		\item $f$ is $\delta$-computable in every network class $\mathcal C^s_{\leq N}$.
	\end{enumerate}
\end{corollary}
\begin{proof}
We only need to prove  that $(ii) \Rightarrow (i)$.
For that, we refine the argument
	in the impossibility proof in \Cref{thm:computability}.
We consider two $\Omega$-vectors of length $n$ and $m$,
	with the same frequency functions
	and continue the proof by replacing
	computability in the class of strongly connected networks
	with computability in the sub-network class~$\mathcal{C}^s_{\leq N}$
	with~$N = \max(n,m)$.
\end{proof}

When the exact size of the network is known,
	deducing multiplicities from frequencies is straightforward,
	and computing frequency-based functions
	thus allows  for computing multiset-based functions.
Hence, knowing the size of the network
	considerably increases the computational power
	in the case of output port awareness,
	symmetric communications, or outdegree awareness,
	while leaving it unchanged
	in the simple broadcast model~\cite{BoldiV:sirocco:1997}.

Any function which is computable in a network class closed under graph isomorphisms
	is necessarily invariant under permutation, i.e., is a multiset-based function.
We thus obtain the following computability result when the network size is known.

\begin{corollary}\label{cor:n}
Let $(X,\delta)$ be a metric space.
In any of the three communication models with either output port awareness, symmetric communications, 
	or outdegree awareness, and for any function $f: \bigcup_{n \in \IN_{>0}} \Omega^n \to X$,
	the following assertions are equivalent:
	\begin{enumerate}[label=(\roman*)]
		\item $f$ is multiset-based;
		\item $f$ is $\delta$-computable in every network class ${\mathcal C}^s_n$.
	\end{enumerate}
\end{corollary}

\subsection{Computing with leaders}

We now study the impact of having leaders on the computational power
	in a static network
	with either output port awareness, symmetric communications, or outdegree awareness.
In the case of a unique leader,
	its fibre is of cardinality one,
	and so the linearity coefficient $k$ in \cref{eq:cardfibre}
	is equal to one.
Hence, agents compute the cardinality of each fibre,
	and our algorithm thus computes any function that is multiset-based.
	
\begin{corollary}\label{cor:leader}
Let $(X,\delta)$ be a metric space.
In any of the three communication models with either output port awareness, symmetric communications, 
	or outdegree awareness, and for any function $f: \bigcup_{n \in \IN_{>0}} \Omega^n \to X$,
	the following assertions are equivalent:
	\begin{enumerate}[label=(\roman*)]
		\item $f$ is multiset-based;
		\item $f$ is $\delta$-computable in the class of static strongly connected networks with one leader.
	\end{enumerate}
\end{corollary}

Observe that this result can be easily extended to the case of $\ell$ leaders
	if~$\ell$ is known of all agents:
	the vertices in a graph~$G$ corresponding to the leaders
	collapse onto some subset~$L_B$  of vertices in the minimum base~$B$ of~$G$.
The cardinality of each fibre $\varphi^{-1}(i)$ is then given by:
	\begin{equation}\label{eq:leaders}
	\left | \varphi^{-1} (i) \right | = \frac{\ell}{\sum_{j\in L_B} z_j}  \, z_i ,
	\end{equation}
	where $\vec{z}$ is the positive integer vector
	whose all entries are coprime and such that~$\ker \mat{M} = \IR \, \vec{z}$.
Hence, our algorithm, together with \cref{eq:leaders},
	allows each agent to compute any multiset-based function.

\section{Dynamic networks}

We now investigate how to compute a frequency-based function
	with symmetric communications or outdegree awareness
	in dynamic networks with a finite dynamic diameter.
In the case of symmetric communications,
	a remarkable recent algorithm due Di Luna and Viglietta~\cite{LunaV:focs:2022,LunaV:disc:2023}
	exactly computes any frequency-based function
	in linear time in the dynamic diameter of the network,
	solving an important open question of computability.
In particular, for a dynamic graph that is strongly connected in each round,
	their algorithm operates in linear time in the size of the network.
Unfortunately, this algorithm is not self-stabilizing
	and does not even tolerate asynchronous starts.
Moreover, this algorithm is based on the construction, by each agent,
	of an infinite \emph{history tree},
	and so uses an infinite number of states
	and an infinite bandwidth in each of its executions.

In this section, we propose to develop another method
	which consists in using consensus algorithms
	derived from statistical physics --
	namely, the \emph{Metropolis} and the \emph{Push-Sum} algorithms.
These algorithms only achieve asymptotic convergence
	and their temporal complexity is non-linear.
However, both tolerate asynchronous starts
	and use no persistent memory.

The Metropolis algorithm
	computes the average of initial values
	in the class of \emph{symmetric} networks
	with a finite dynamic diameter
	and in the communication model of outdegree awareness,
	even under asynchronous starts.
Its convergence rate has been showed to be quadratic~\cite{Charron-Bost:infcomp:2022}
	in the case of a dynamic network
	that is strongly connected in each round.\footnote{%
	Using the \emph{Lazy Metropolis} algorithm~\cite{Olshevsky:siamco:2017,NedicOR:pieee:2018},
	this result can be extended
	to the case of (symmetric) networks
	with a finite dynamic diameter.}
A variant for the simple model of symmetric communications
	(without assuming outdegree awareness)
	has been proposed~\cite{DBLP:phd/hal/Lambein20,Charron-BostL:sand:2022},
	but its temporal complexity is in $\bigO{n^4}$.
	
In the case of outdegree awareness, we develop  another approach based on  the \emph{Push-Sum} algorithm.
The algorithm was introduced in~\cite{KempeDG:focs:2003}, where its correctness  was shown 
	in a probabilistic communication model  with  pairwise communications in the fully-connected graph.
This result was then extended to arbitrary strongly connected graphs in~\cite{BenezitBTTV:isit:2010}.
Further, Nedi\`c et al.~\cite{NedicOR:pieee:2018} proved the correctness of Push-Sum in any dynamic network 
	with a finite diameter.
Below, we give a  self-contained and streamlined convergence proof, and then describe how Push-Sum can be used to compute
	frequency-based functions.

\subsection{The Push-Sum algorithm and the quot-sum function}

The Push-Sum algorithm proceeds as follows:
	each agent $i$ maintains three variables $x_i$, $y_i$, and~$z_i$.
The two variables $y_i$ and $z_i$ are initialized respectively to
	$v_i \in \IR$ and $w_i \in \IR_{>0}$,
	and they are updated as follows:
	\begin{align}
		y_i(t) &= \sum_{k \in \In_i(t) }\frac{y_k(t -1 )}{d_k^-(t)}
		\label{ps:x_update}\\
		z_i(t ) &= \sum_{k \in \In_i(t)}\frac{z_k (t -1)}{d_k^-(t)}
		\label{ps:y_update} \enspace.
	\end{align}
The variable $x_i$ is initialized to $v_i/w_i$ and set to $x_i  = y_i / z_i $ at the end of each round.
Observe that by the very definition of its update rules, the Push-Sum algorithm requires output port awareness.

The main result of this section is that, under the assumption of a network with a finite dynamic diameter,
	the Push-Sum algorithm computes the \emph{quot-sum function} defined by
	 \begin{equation*}
		 \begin{aligned}
			 \qs : & \bigcup_{n \in \IN_{>0}}( \IR\times \IR_{>0})^n & \to     & \: \IR \\
						 & \big((v_1,w_1), \cdots, (v_n,w_n)\big)          & \mapsto & \: \frac{\sum_{k \in [n]} v_k}{\sum_{k\in [n]} w_k} \enspace.
		 \end{aligned}
	 \end{equation*}
In other words, for each agent $i$,
	the quotient $x_i(t)/y_i(t)$ asymptotically converges
	to the quot-sum of the initial values.

\subsection{Preliminaries}\label{subsec:pushsum:preliminaries}

We first introduce some notation.
Let $n$ be a positive integer,
	$\vec{v} \in \IR^n $ a real vector,
	and $\mat{A} \in \IR ^n \times \IR ^n $
	a real square matrix, both of size $n$.
The vector~$\vec{v}$ or the matrix~$\mat{A}$
	is said to be \emph{non-negative} (resp. \emph{positive})
	if all its entries are non-negative (resp. positive).
The \emph{graph associated to} a non-negative matrix~$\mat{A}$
	is the directed graph $G_{\mat{A}}= ([n], E_{\mat{A}})$,
	where $E_{\mat{A}}$ is the set of edges defined as
	\begin{equation*}
		E_{\mat{A}} \eqdef \{ (j,i) \in [n]^2 : A_{i,j}  > 0 \} .
	\end{equation*}
A vector is \emph{stochastic}
	if it is non-negative
	and its entries sum to~$1$;
	a \emph{matrix} is in turn (row-)\emph{stochastic}
	if each of its rows is a stochastic vector;
	correspondingly, a matrix is \emph{column-stochastic}
	if each of its \emph{columns} is a stochastic vector.
Importantly, the sum of entries of a vector
is left invariant by any column-stochastic matrix $\mat{A}$ --
	namely, $\sum_{i \in [ n ]} (\mat{A} \, \vec{v})_i = \sum_{i \in [ n ]}  v_i $.

Let $\mat{A} = (\mat{A}(t))_{t\in \IN_{>0}}$ be a sequence
	of non-negative square matrices of size $n$.
The entry of $\mat{A}(t)$ at the $i$-th row and $j$-th column is denoted by $A_{i,j}(t)$.
For any positive real number~$\alpha$,
	the matrix $\mat{A}(t)$ is said to be \emph{$\alpha$-safe}
	if all its positive entries lie in the interval~$[\alpha,+\infty)$.
The graph associated to the matrix $\mat{A}(t)$ is denoted $\dG_A(t)$,
	and the sequence~$\dG_A(1), \dG_A(2), \cdots$ 
	forms a \emph{dynamic graph} $\dG_{\mat{A}}$
	over the set of nodes $[ n ]$.
If $t, t' \in \IN_{>0}$, with $t' \geq t$,
	we let $\mat{A}(t':t)$
	denote the backward product~$\mat{A}(t') \times \cdots \times \mat{A}(t)$.
	Hence, the graph associated to the matrix~$\mat{A}(t':t)$
	is the (forward) product
	$\dG(t:t') = \dG(t) \circ \cdots \circ \dG(t')$.

\begin{lemma}\label{lem:vs}
	Let $(\mat{A}(t))_{t\in \IN_{>0}}$
	be a sequence of $\alpha$-safe column-stochastic matrices of size~$n$
	with positive diagonal entries,
	and let $\vec{v} \in \IR^n$ be a non-negative vector.
If the associated dynamic graph $\dG_{\mat{A}}$
	has a finite dynamic diameter~$D$,
	then for all~$i \in [ n ] $ and all~$t \geq D$, we have
	\begin{equation*}
		\alpha^{D}\sum_{k \in [n]} v_k
		\leq v_i(t)
		\leq \sum_{k \in [n]} v_k
	\end{equation*}
	where $\vec{v}(t) \eqdef \mat{A}(t:1)\, \vec{v}$.
\end{lemma}
\begin{proof}
By induction, we easily check that, for any $t\in \IN$,
	all the entries of $\vec{v}(t)$ are positive.
Moreover, the sum of the entries in $\vec{v}(t)$ is invariant with~$t$.
It follows that, for every integer $t \in \IN$, we have
\begin{equation*}
	v_i(t) \leq \sum_{k \in [ n ]} v_k (t) = \sum_{k \in [ n ]} v_k  .
\end{equation*}

For the lower bound,
	we start by observing that
	since all the positive entries
	of the matrices~$\mat{A}(1), \mat{A}(2), \cdots$
	are at least equal to $\alpha$,
	the positive entries of every product of $t$ matrices in this sequence
	are at least equal to $\alpha^{t}$.
In particular, if $t \geq D$,
	then all the positive entries of the matrix $A(t:t-D+1)$
	are at least equal to $\alpha^{D}$.
Moreover, the graph associated to the product matrix~$\mat{A}(t:t-D+1)$
	is the graph $\dG(t-D+1 : t)$
	equal to the complete graph,
	since~$D$ is the dynamic diameter of $\dG$.
Therefore, all the entries of the matrix $\mat{A}(t:t-D+1)$ are positive,
	and thus at least equal to $\alpha^{\,D}$.

Pick $t \geq D$.
We have $\vec{v}(t) = \mat{A}(t:t-D+1) \, \vec{v}(t- D)$, and thus
	\begin{equation*}
		v_i(t) = \sum_{k\in [n]} A_{i,k} (t:t-D+1) v_k (t- D) \geq \alpha^{D}   \sum_{k\in [n]}  v_k (t-D)  .
	\end{equation*}
Since the sum of the entries in $\vec{v}(t)$ is constant, we obtain
	$v_i(t)  \geq  \alpha^{D}  \sum_{k\in [n]}  v_k $.
\end{proof}

\subsection{Push-Sum for computing the quot-sum function}

We are now in position to prove that the Push-Sum algorithm computes the quot-sum of initial values.

\begin{theorem}\label{thm:pushsum}
The Push-Sum algorithm computes the quot-sum function
	in the class of networks with a finite dynamic diameter.
More precisely, in any execution of Push-Sum
	with a network of dynamic diameter~$D$,
	all the output variables are within $\varepsilon$
	of the quot-sum of the initial values
	in $\bigO*{n^{2D} \log\frac{1}{\varepsilon}}$ rounds.
\end{theorem}

\begin{proof}
We first consider an execution of the Push-Sum algorithm
	in a dynamic network $\dG$ with $n$ agents
	and a finite dynamic diameter~$D$,
	and synchronous starts at round one for all agents.
Observe that if $A(t)$ is the square matrix defined from the directed graph $\dG(t)$ by 
	\begin{equation*}
		A_{i,j}(t) =
		\begin{cases*}
			1/\outd_j (t) & if $(j,i) \in E(t)$ \\
			0             & otherwise
		\end{cases*}
	\end{equation*}
	where $\outd_j (t) $ denotes the outdegree of $j$ in $\dG(t)$,
	then $A(t)$ corresponds to the update rules for the variables
	$y_i$ and $z_i$ at round~$t$, namely
	\begin{equation*}
		\vec{y}(t) = \mat{A}(t) \, \vec{y}(t-1) \: \text{and} \:
		\vec{z}(t) = \mat{A}(t) \, \vec{z}(t-1) .
	\end{equation*}
Each matrix $\mat{A}(t)$ is column-stochastic and $\frac{1}{n}$-safe.
\Cref{lem:vs} shows that for all $i \in [ n ]$ and $t \geq D$,
	\begin{equation}\label{eq:y}
	n^{-D} \sum_{k \in [ n ]} w_k  \leq y_i(t) \leq \sum_{k \in [ n ]} w_k  .
	\end{equation}
The vector $\vec{z}(t)$ is positive, and thus
	$\vec{x}(t) = [\diag(\vec{z}(t))]^{-1} \vec{y}(t)$.
It follows that
	\begin{equation*}
		\vec{x}(t) =  \mat{B}(t) \, \vec{x}(t -1)
	\end{equation*}
	where $\mat{B}(t) \eqdef {[\diag(\vec{z}(t))]}^{-1} \mat{A}(t) \, \diag(\vec{z}(t-1))$.

We easily check that
	all the entries of $\mat{B}(t)$ are non-negative,
	and $\mat{B}(t)$ is a stochastic matrix
	with a positive diagonal.
Its associated graph is the same as $\mat{A}(t)$'s, namely $\dG(t)$.

The next step of the proof
	consists in proving that the product matrix $\mat{B}(t:1)$
	converges to a rank one matrix.%
	\footnote{This convergence result can be directly deduced
		from a theorem~\cite[Theorem~3]{caoMA:siamjco:2008}
		established by Cao, Morse, and Anderson for dynamic graphs
		that are rooted with bounded delay.
	The proof that we develop here
		in the particular case of a finite dynamic diameter
		is simpler and provides a better bound on the convergence rate.}
Let us first observe that
	Dobrushin's ergodic coefficient~\cite[eq.~(1.5)]{Dobrushin:tpia:1956}
	of a stochastic matrix $\mat{P}$,
	defined by
	\begin{equation*}
		\delta(\mat{P}) \eqdef 1 - \min_{i \neq j} \sum_{k \in [n]} \min (P_{i,k}, P_{j,k})
	\end{equation*}
	lies in the range $[0,1]$ and satisfies the inequality
	\begin{equation*}\delta (\mat{P}) \leq 1 - n \alpha \end{equation*}
	when $\mat{P}$ is $\alpha$-safe and its associated graph is fully-connected.
A result by Seneta~\cite{Seneta:aap:1979}
	combined with a straightforward argument of convex duality
	shows that for any stochastic matrix~$\mat{P}$,
	the ergodic coefficient $\delta(\mat{P})$ coincides with the matrix seminorm
	\begin{equation*}
		\sup_{\delta(\vec{v}) > 0} \frac{\delta (\mat{P} \, \vec{v})}{\delta(\vec{v})}
	\end{equation*}
	associated to the seminorm on $\IR^n$ defined by
	$\delta(\vec{v})= \max_{i \in [n]} v_i -  \min_{i \in [n]} v_i$.
Consequently,
	the ergodic coefficient~$\delta$ is a matrix seminorm,
	and so is sub-multiplicative.

The sequence $(\mat{A}(t))_{t\in \IN_{>0}}$ is $\frac{1}{n}$-safe,
	and the inequalities in \cref{eq:y}
	show that every matrix product
\begin{equation*}
	\mat{B}(t+D-1:t) = {[\diag(\vec{z}(t+D-1))]}^{-1} \mat{A}(t+D-1:t) \, \diag(\vec{z}(t-1))
\end{equation*}
	is $n^{-2D} $-safe.
It follows that
	\begin{equation*}
		\delta \left(\mat{B}(t:1)\right) \leq \left (1 - n^{-2D} \right)^{\lfloor t/D\rfloor} \, .
	\end{equation*}
Because of the inequality $\log(1-a) \leq -a$,
	valid whenever $a \geq 0$,
	we obtain that if $t \geq D\, n^{2D} \log \big(\frac{1}{\varepsilon}\big)$,
	then $\delta \left(\mat{B}(t:1)\right) \leq\varepsilon$
	and $\delta \left(x(t)\right) \leq\varepsilon \, \delta \left(x(0)\right)$.

This shows that $\lim_{t\to \infty} x^+(t) -x^-(t) = 0$,
	where $x^-(t) \eqdef \min_{i \in [n]} x_i(t)$
	and $x^+(t) \eqdef \max_{i \in [n]} x_i(t)$.
Since each matrix $\mat{B}(t)$ is stochastic,
	the sequences $\big(x^+(t)\big)_{t \in \IN} $ and $\big(x^-(t)\big)_{t\in\IN} $
	are non-increasing and non-decreasing, respectively.
Hence, the two sequences converge to the same limit,
	that we denote $x^*$, and all the sequences $\big(x_i(t)\big)_{t\in\IN} $
	also converge to $x^*$.
The convergence rate follows from the above.

Every sequence $(z_i(t))_{t\in\IN}$ is bounded (see \cref{eq:y}),
	and hence $\lim_{t\to \infty} y_i(t) - x^* z_i(t) = 0$ .
Since the sum of the entries in $\vec{y}(t)$ and $\vec{z} (t)$ are constant,
	this implies that $\sum_{k \in [ n ]} v_k = x^* \sum_{k \in [ n ]} w_k $.	
\end{proof}

Push-Sum is not a self-stabilizing algorithm
	(initializations of the variables $y_i$ and $z_i$ cannot be arbitrary),
	but it tolerates asynchronous starts.
Clearly, an execution with the dynamic graph~$\dG$ and the agents $i$ starting at rounds $s_i$
is similar to the execution
	 where all the agents start at round one and with the dynamic graph
	 $\widetilde{\dG}$ with $n$ vertices and the set of edges defined by:
	 \begin{equation*}\widetilde{E}_t = \{ (i,j) \in E_t : i = j \, \vee \, t \geq \max(s_i,s_j)  \} .\end{equation*}
Observe that if $\dG$ has a finite dynamic diameter $D$, then 
	$ \max (s_i) + D$ is an upper bound on $\widetilde{\dG}$'s dynamic diameter.

\subsection{Push-Sum for computing a frequency-based function}\label{sec:push-sum-func}

A Push-Sum based algorithm for computing the frequency function
	$ \nu : v  \in \Omega^n \rightarrow  \nu_v \in  \IQ^{\Omega}$
	can be easily derived from \Cref{thm:pushsum} (see Algorithm~\ref{algo:pushsum}).
The variables $x_i$, $y_i$, and $z_i$ are now three arrays of dynamic size,
	each of which initially contains only one variable,
	indexed by the initial value of vertex~$i$ and equal to one.
As soon as the vertex~$i$ becomes aware
	that some value~$\omega \in \Omega$ is initially present in the network,
	upon the first receipt of some variables indexed by $\omega$,
	it appends three variables to $x_i$, $y_i$, and $z_i$
	all indexed by $\omega$, and respectively initialized to $0$, $0$, and $1$.
Then, vertex~$i$ starts to run the Push-Sum algorithm
	with the variables $y_i[\omega]$ and $z_i[\omega]$.

\begin{algorithm}
	\caption{The Push-Sum algorithm for computing the frequency function}\label{algo:pushsum}
	\Input{$v_i \in \Omega$}
	\Initially{
		$x_i[v_i] \gets 0$,
		$y_i[v_i] \gets 1$,
		$z_i[v_i] \gets 1$
	}
	\EachRound{
		\textbf{send} $\langle y_i, z_i , \outd_i \rangle$ to all\;
		\textbf{receive}
			$\langle y_{j_1}, z_{j_1}, \outd_{j_1} \rangle,
			\cdots,
			\langle y_{j_{\ell}}, z_{j_{\ell}}, \outd_{j_{\ell}} \rangle$
			from the in-neighbors
			\Comment*{$\ell$ in-neighbors}
		\For{$\omega$ appearing in the support
			of any vector~$y_i, y_{j_1}, \ldots, y_{j_{\ell}}$}{
			\For{$k = 1, \ldots, \ell$}{
				\If{$\omega$ is not in the support of~$y_{j_k}$}{
					$y_{j_k}[\omega] \gets 0$,
					$z_{j_k}[\omega] \gets 1$
				}
			}
			\If{$\omega$ is not in the support of~$y_i$}{
				$x_i[\omega] \gets 0$,
				$y_i[\omega] \gets 0$,
				$z_i[\omega] \gets 1$
			}
		}
		\For{$\omega$ in the support of~$y_i$}{
			$y_i[\omega] \gets \sum_{k=1}^{\ell} y_{j_k}[\omega] / \outd_{j_k}$ \;
			$z_i[\omega] \gets \sum_{k=1}^{\ell} z_{j_k}[\omega] / \outd_{j_k}$ \;
			$x_i[\omega] \gets y_i[\omega] / z_i[\omega]$ \;
		}
	}
\end{algorithm}

For each value $\omega \in \Omega$,
	the execution of this algorithm
	corresponds to one instance of the Push-Sum algorithm
	initiated by the vertices whose initial value is $\omega$.
Since Push-Sum tolerates asynchronous starts,
	the frequency of the value~$\omega$
	is asymptotically computed
	in the corresponding variable in the $x_i$ array
	if $\omega$ is the initial value of some agent,
	or is equal to zero
	if $\omega$ is not initially present in the network.

Observe that the frequency of the value~$\omega$
	in the vector $\vec{v} \in \Omega^n$~%
	-- namely, $\nu_{\vec{v}}(\omega)$ --
	is a rational number in the \emph{finite} set
	\begin{equation*}
	\IQ_n = \left\{
	\frac{p}{q} \in \IQ :
		p \in \IN \wedge q \in \IN_{>0} \wedge 0 \leq p \leq q \leq n
	\right\} \enspace.
\end{equation*}
If the agent~$i$ knows an upper bound $N$ on the network size,
	then $i$ can determine the set $\IQ_N \supseteq \IQ_n$
	and, at each round,
	it can compute the nearest rational number to $x_i[\omega](t)$ in $\IQ_N$.
\Cref{thm:pushsum} shows that
	these rational numbers are eventually all equal to $\nu_{\vec{v}}(\omega)$.
In this way, agent~$i$ computes the frequency function
	in any vector $\vec{v} \in \Omega^n$ in finite time,
	and hence the value of $f(\vec{v})$,
	if $f$ is a frequency-based function.
Since two different numbers in $\IQ_N$ are at distance at least $1/N^2$,
	we obtain that the stabilization time of
	the algorithm is in $\bigO{n^{2D} \log N}$.

Let $\mathcal C_{ \leq N}$ denote the class of \emph{dynamic} networks with at most $N$ vertices and a finite dynamic diameter.
Combined with the impossibility result in Corollary~\ref{cor:boundn}, we obtain the following characterization 
	of computable functions in dynamic anonymous networks with a finite dynamic diameter, when a bound
	on the network size is known.
\begin{corollary}\label{cor:boundndyn}
Let $(X,\delta)$ be a metric space.
With outdegree awareness, for any function $f: \bigcup_{n \in \IN_{>0}} \Omega^n \rightarrow X$,
	the following assertions are equivalent:
	\begin{enumerate}[label=(\roman*)]
		\item $f$ is frequency-based;
		\item $f$ is $\delta$-computable in every network class $\mathcal C_{\leq N}$.
	\end{enumerate}
\end{corollary}

Since any function which is computable in a network class closed under graph isomorphisms is necessarily a multiset-based function,
	we derive the following computability result when the network size is known.
\begin{corollary}\label{cor:ndyn}
Let $(X,\delta)$ be a metric space.
With outdegree awareness, for any function $f: \bigcup_{n \in \IN_{>0}} \Omega^n \rightarrow X$,
	the following assertions are equivalent:
	\begin{enumerate}[label=(\roman*)]
		\item $f$ is multiset-based;
		\item $f$ is $\delta$-computable in every network class $\mathcal C_{n}$.
	\end{enumerate}
\end{corollary}	
	
We now study how to use the Push-Sum algorithm
	for computing a function
	when no bound on the network size is available.
For that, we first observe that at each round~$t$,
	the positive rational numbers $x_i[\omega](t)$
	may not correspond to a frequency vector,
	since their sum may be different from one.
This is why each agent~$i$
	maintains an additional variable~$\widetilde{\vec{x}_i}$, set to
	\begin{equation*}
		\widetilde{x_i}[\omega] = \frac{x_i[\omega]}{\sum_{\omega'} x_i[\omega']}
	\end{equation*}
	to form a frequency function.
Then, agent~$i$
	can easily construct a vector $\langle \widetilde{\vec{x}_i} \rangle$ on $\Omega$
	that is $ \widetilde{\vec{x}_i}$-frequenced,
	i.e., whose frequency function is $ \widetilde{\vec{x}_i}$.

This leads us to introduce
	the notion of \emph{$\delta$-continuity in frequency}
	for a frequency-based function~$f: \bigcup_{n \in \IN_{>0}} \Omega^n \rightarrow X$.
If $\vec{v}(1), \vec{v}(2), \cdots$
	is a sequence of vectors of arbitrary size
	such that, for every value $\omega\in \Omega$,
	the sequence of frequencies~%
	$\nu_{\vec{v}(1)}[\omega], \nu_{\vec{v}(2)}[\omega], \cdots$
	converges to some limit value~$\nu^* [\omega]$
	and those limit values form a frequency function $\nu^*$,
	then the sequence $f(\vec{v}(1)), f(\vec{v}(2)), \cdots$
	converges in $(X, \delta)$ to $f\big( \langle \nu^* \rangle \big)$.
	
As an example,
	the average function is continuous in frequency
	with the classical metric on $\IR$.
Other examples are provided by the \emph{threshold frequency predicates} $\Phi_r^{\omega} : \bigcup_{n \in \IN_{>0}} \Omega^n \rightarrow \{0,1\}$, 
	where $r$ is a real number in $[0,1]$ and $\omega\in \Omega$, defined by
	\begin{equation*}
		\Phi_r^{\omega} (\vec{v}) \eqdef
		\begin{cases*}
			1 & if $\nu_{\vec{v}}(\omega) \geq r$ \\
			0 & otherwise.
		\end{cases*}
	\end{equation*}
Indeed, the function $ \Phi_r^{\omega} $
	is continuous in frequency
	with the discrete metric on $\{0,1\}$
	if and only if $r$ is \emph{irrational}.

Algorithm~\ref{algo:pushsum},
	complemented with the variables $\widetilde{\vec{x}_i}$,
	then $\delta$-computes any function~$f$
	that is $\delta$-continuous in frequency:
	the variables $f \big( \langle  \widetilde{\vec{x}_i} \rangle \big)$
	all tend to $f(\vec{v})$ (in the sense of the metric~$\delta$).

\begin{corollary}\label{cor:continuity}
Let $(X,\delta)$ be a metric space.
In the communication model with outdegree awareness,
	every frequency-based function
	$f: \bigcup_{n \in \IN_{>0}} \Omega^n \rightarrow X$
	that is $\delta$-continuous in frequency
	is $\delta$-computable
	in the class of dynamic networks of finite dynamic diameter.
\end{corollary}

\subsection{Computing with leaders}\label{sec:push-sum-leaders}

If there is a set of~$\ell \geq 1$ leaders in the network,
	with~$\ell$ known to all agents,
	a slight variant of the Push-Sum algorithm allows each agent~$i$ to compute
	any multiset-based function:
	its code is unchanged
	except if it is not a leader
	in which case its variables $z_i[\omega]$
	are initially set to zero instead of one
	(cf. lines~3, 10, and~12 in Algorithm~\ref{algo:pushsum};
	it is then possible, on line~16,
	that~$x_i[\omega]$ be equal to~$\infty$,
	but only for finitely many rounds).
The variable~$\ell x_i[\omega]$ tends to $\omega$'s multiplicity, which is thus asymptotically computable.
The frequency-based condition can thus be replaced by the multiset-based condition in \Cref{cor:boundndyn,cor:continuity}.

\section{Concluding Remarks}


In this paper, we have presented
	a panorama of function computability by anonymous networks,
	both static and dynamic,
	for the various communication models
	that are typically considered
	when studying message-passing distributed systems.

Three classes of functions stand out:
	\emph{set-based} functions,
	whose output is determined by the \emph{set} of values
	appearing in the input vector;
	\emph{multiset-based} -- or symmetric -- functions,
	which are determined by the multiset of values;
	and the intermediate class of \emph{frequency-based functions},
	whose values may depend on the relative frequency of the input values,
	but not on their multiplicities.

A fundamental result states that
	in anonymous networks communicating through a simple \emph{local broadcast} primitive~%
	-- where an agent has no knowledge or control
	of the recipients of its messages --
	only set-based functions can be computed,
	both in the dynamic and in the static case.
This holds even if we assume global symmetry breaking
	in the form of one or several agents being designated as leaders.
Conversely, a simple flooding algorithm
	easily allows all agents to recover
	the set of all input values in finite time,
	and thus to compute any set-based function.

In hope of computing a larger class of functions,
	we have described three ways
	of augmenting this simple communication model:
	we can work under the assumption that communication links are symmetric;
	we can assume that senders are aware, ahead of emission,
	of how many other agents will receive each of their messages;
	or, in the case of a static network,
	we can assume that an agent
	may individually address each of its neighbors.

Under all three models for static networks,
	and under the former two for dynamic networks,
	the class of computable functions is almost characterized
	as that of the frequency-based functions,
	up to some restrictions in the dynamic case, discussed hereafter.
Moreover, under any of these models,
	breaking the symmetry by introducing one or several leaders
	allows for recovering the full multiset of input values,
	and thus for computing multiset-based functions,
	as does providing the agents with the size of the network,
	but \emph{not} with a bound over the size of the network.
For static networks, these results are collected in \Cref{fig:tab},
	and for dynamic networks they are presented in \Cref{fig:tabdyn}.

There are slight limitations to the above picture
	that we expect further works to address.
First, our Push-Sum-based method for computing frequency-based functions
	only works if a bound over the number of agents is known by the agents,
	in order to turn an approximate result into an exact one.
Otherwise, we must restrict ourselves
	to functions that are additionally \emph{continuous in frequency}.
Can we lift these restrictions
	and recover the same computability statement
	for the static and the dynamic case?
What exactly characterizes continuity in frequency?

Another consideration, still regarding the dynamic case,
	concerns the connectivity assumption.
The algorithms that we consider
	have not, in general, been shown to work
	under the relaxed assumption of a network
	that, while never becoming permanently split,
	do not have a finite dynamic diameter --
	asides from the Metropolis-based family of algorithms.
The convergence of the latter,
	for a symmetric communication model,
	results from Moreau's remarkable theorem~\cite[Theorem 1]{Moreau:itac:2005},
	which ensures the convergence of a wide family of algorithms.
This weaker connectivity assumption
	is often considered when studying natural systems
	through a distributed lens~\cite{AngluinADFP:dc:2006,AngluinAER:dc:2007,chazelle2011sjco,Chazelle:cacm:2012}.
Which of our computability results continue to hold in this case?
The recent algorithm designed by Di Luna and Viglietta~\cite{LunaV:focs:2022,LunaV:disc:2023}
	for the case of symmetric networks,
	could conceivably work;
	what, however, can be said in the outdegree awareness model,
	where Moreau's theorem does not apply?

Finally, what of self-stabilizing computation?
Here again, neither Di Luna and Viglietta's algorithm, nor Push-Sum, continue to work.
Can either of them be fixed?
If not, what can be said of self-stabilizing computation
	over dynamic networks,
	under the different communication models that we consider?

\begin{acks}
	We would like to thank Jérémie Chalopin,
		Louis de Monterno, Alex Olshevsky,
		and Michaël Thomazo
		for useful discussions related to this work.
\end{acks}

\bibliographystyle{plainurl}
\bibliography{references}

\end{document}